\documentclass[conference]{IEEEtran}
\IEEEoverridecommandlockouts
\usepackage{cite}
\usepackage{amsmath,amssymb,amsfonts}
\usepackage{algorithmic}
\usepackage{graphicx}
\usepackage{svg}
\usepackage{bm}
\usepackage{textcomp}
\usepackage{xcolor}
\usepackage{hyperref}
\usepackage{enumitem}
\def\BibTeX{{\rm B\kern-.05em{\sc i\kern-.025em b}\kern-.08em
    T\kern-.1667em\lower.7ex\hbox{E}\kern-.125emX}}
    
\begin{document}

\title{CREDENCE: Counterfactual Explanations for Document Ranking}
%\title{CREDENCE: Counterfactual Explanations of Rankings}
% \title{CREDENCE: Creating Document Ranking Explanations Counterfactually}

%%% To align author names in the center
\makeatletter
\newcommand{\linebreakand}{%
  \end{@IEEEauthorhalign}
  \hfill\mbox{}\par
  \mbox{}\hfill\begin{@IEEEauthorhalign}
}
\makeatother

% Macro for the name of the tool
%\def\keyword#1{\setformath{$\sf{#1}$}}
\def\keyword#1{$\sf{#1}$}
\newcommand{\toolname}{\keyword{CREDENCE}}

% Authors
\author{
\IEEEauthorblockN{Joel Rorseth}
\IEEEauthorblockA{
\textit{University of Waterloo}\\
jerorset@uwaterloo.ca}
\and
\IEEEauthorblockN{Parke Godfrey}
\IEEEauthorblockA{
\textit{York University}\\
godfrey@yorku.ca}
\and
\IEEEauthorblockN{Lukasz Golab}
\IEEEauthorblockA{
\textit{University of Waterloo}\\
lgolab@uwaterloo.ca}
\linebreakand
\IEEEauthorblockN{Mehdi Kargar}
\IEEEauthorblockA{
\textit{Toronto Metropolitan University}\\
kargar@ryerson.ca}
\and
\IEEEauthorblockN{Divesh Srivastava}
\IEEEauthorblockA{
\textit{AT\&T Chief Data Office}\\
divesh@research.att.com}
\and
\IEEEauthorblockN{Jaroslaw Szlichta}
\IEEEauthorblockA{
\textit{York University}\\
szlichta@yorku.ca}
}

\maketitle

\begin{abstract}

Towards better explainability in the field of information retrieval, we present \toolname{}, an interactive tool capable of generating counterfactual explanations for document rankers. Embracing the unique properties of the ranking problem, we present counterfactual explanations in terms of document perturbations, query perturbations, and even other documents. Additionally, users may build and test their own perturbations, and extract insights about their query, documents, and ranker.

\end{abstract}

% \begin{IEEEkeywords}
% Explainability, Counterfactual, Ranking, XAI
% \end{IEEEkeywords}

\section{Introduction}

%Counterfactual explanations have become a popular and
%effective format to explain decisions made by artificial
%intelligence (AI) models. We introduce CREDENCE,
%the first tool allowing users to generate
%counterfactual explanations for document ranking models.

With the rise of deep learning (DL),
significant advances have been made by the data science
community, though often at the cost of increased model
complexity. For many modern DL models, the underlying
decision-making process is nearly unintelligible
for data scientists and users \cite{poem}.
With the growing
adoption of data science in critical domains,
such as medicine and law, \textit{explainability}
has become a priority in many deployment scenarios.
In critical applications, explanations build trust
between models and their users, and enable auditing
that works to ensure regulation adherence, mitigation
of bias, and sufficient justification.

In recent years, researchers have developed a variety
of solutions that support
\textit{explainable artificial intelligence (XAI)},
which combat the increasing complexity that renders
DL models uninterpretable. Among different types
of \textit{local} explanations, which aim to
rationalize individual predictions (decisions),
\textit{counterfactual} explanations
\cite{certa}\cite{xaidb} have emerged
as a popular and pragmatic explanation format to
impart behavioral insight. Generally, counterfactual
explanation methods identify sets of minimal changes
to the features of an input, such that a change is
observed in a model's prediction.

In the field of information retrieval (IR),
complex DL models have been employed for a variety
of tasks, most notably for document ranking.
Naturally, the decision-making and ranking logic
behind document ranking models (rankers)
is often unclear to their users \cite{explranking}.
%In most closely related work, 
Explainability for
document rankers has been
limited to derivatives of
\textit{saliency} explanations
\cite{singh2019exs} \cite{lirme}
\cite{fernando2019study}
\cite{chios2021helping},
which attempt to approximate the relative
importance of model features
(e.g., query or document terms).
To the best of our knowledge, counterfactual
explanations have not yet been adapted for
document rankers. 
To 
%expand the current state of the art, 
fill this gap, we
demonstrate \toolname, the first tool for
\emph{CREating DocumEnt raNking explanations CountErfactually}.%
\footnote{%
    A video is available at
    \url{https://vimeo.com/762787210}\label{video-link}. \\
    %%% $^{,}$
    %%% Wha?? -- Parke
    The tool is available at
    \url{http://lg-research-1.uwaterloo.ca:8091/credence}%
    \label{system-link}.
}

% \footnote{Video available at \url{https://www.youtube.com/watch?v=-xwUQlBAfgo}.
% Tool available at \url{http://lg-research-1.uwaterloo.ca:8091/credence} \label{system-link}}

Our interactive tool produces several types of
counterfactual explanations, which
collectively expose the decision-making
logic behind a ranking model:

\begin{enumerate}[nolistsep,leftmargin=*]
    \item \textbf{Counterfactual Documents.}
    Explore minimal perturbations to a given
    document that lower its rank (towards the bottom of the ranking)
    beyond some threshold.

    \item \textbf{Counterfactual Queries.}
    Explore minimal perturbations to a search query that
    raise the rank of a given document (towards the top of the ranking).

    \item \textbf{Instance-Based Counterfactual Documents.}
    For a given relevant document, discover similar
    documents that were deemed non-relevant.

    \item \textbf{Build-Your-Own Counterfactual Documents.}
    Interactively edit a given ranked document,
    then compare the resulting ranking against
    the original.
\end{enumerate}

\section{System Description}

\subsection{Preliminaries}

In the document ranking problem, a user poses a search
query $q$ to a ranking model $M$. Given a
set of indexed documents $D$ (i.e., the corpus),
the ranking model $M$ is tasked with producing
a ranking (i.e., an ordered list of
documents) $\boldsymbol{D}^M$ such that,
when treated like a set,
$\boldsymbol{D}^M \subseteq D$.
Naturally, $q$, $D$, and $M$ jointly
contextualize the definition of
$\boldsymbol{D}^M$.
In practice, it is often the case that
$|\boldsymbol{D}^M| \ll |D|$, since many rankers need only to identify and rank the top-$k$ \textit{relevant}
documents (i.e., $|\boldsymbol{D}^M| = k$ for some
parameter $k$).

%Thus, for notational convenience, 
Let $R(q, d, D, M)$ denote the ranking function
representing a ranking model $M$. $R$ returns
the rank $r \in [1, |D|]$ assigned by $M$,
corresponding to the predicted relevance of a
document $d \in D$ to a search query $q$.
$R$ and $M$ are defined generally, such that the
ranker (e.g., a machine learning model) is considered a \textit{black box}. However,
we assume that $R$ assesses rank using
only the body of each document. In future
work, we plan to explain ranking models
that support richer sets of features
(e.g., user preferences).

\subsection{System Architecture}

\toolname{} is an interactive web application built
with the React framework, along with other
JavaScript libraries, such as Material UI
to render user interface components. The
backend is implemented in Python 3.9.14, and
ultimately runs as an ASGI web server (via
Uvicorn). Our server takes the form of a REST
API, built using the FastAPI framework,
which exposes endpoints to retrieve all
data displayed in the web application.
Both applications are hosted on
a server running Ubuntu 22.04, with an
AMD Opteron 6348 Processor, 128 GB of DDR3 RAM,
and GeForce RTX 2080 Ti GPU.

\begin{figure}[t]
\centerline{
    %\includesvg[width=\textwidth]{architecture.svg}
    \includesvg[scale=0.45]{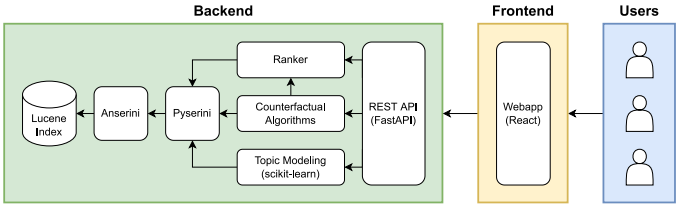}
}

\vskip -0.1in
\caption{The architecture behind the \toolname{} system.}
\vskip -0.13in
\label{architecture}
\end{figure}

The architecture of \toolname{} is  illustrated in Figure \ref{architecture}.
To facilitate all retrieval functionality,
we create a Lucene index using the
Pyserini library
\cite{Lin_etal_SIGIR2021_Pyserini},
which is a Python interface for the
Anserini retrieval toolkit \cite{anserini}.
Although any compatible ranker could be
used to rank documents in our index, we
utilize the monoT5 neural ranker from
the PyGaggle library.%
\footnote{\url{http://pygaggle.ai}.}
%, %TODO Citation needed%
%for the purposes of this demo.
We implement several counterfactual
algorithms, each repeatedly querying
the ranker and index to develop
understanding of the relationships between
documents, search queries, and their rankings.
Also, we offer a topic modeling
module, allowing users to browse
clusters of terms found in selected
documents, for the purpose of discovering
important terms that may influence
relevance. Topic modeling
capabilities are enabled through the
scikit-learn  
%\cite{scikit-learn}
implementation of the Latent Dirichlet
Allocation (LDA) model \cite{lda}.
Using the FastAPI framework, we expose
REST endpoints to perform ranking,
generate counterfactual explanations, and
discover topics.

\begin{figure*}[t]
\vskip -0.1in
\begin{minipage}[t]{0.315\textwidth}
  \includegraphics[width=\linewidth]{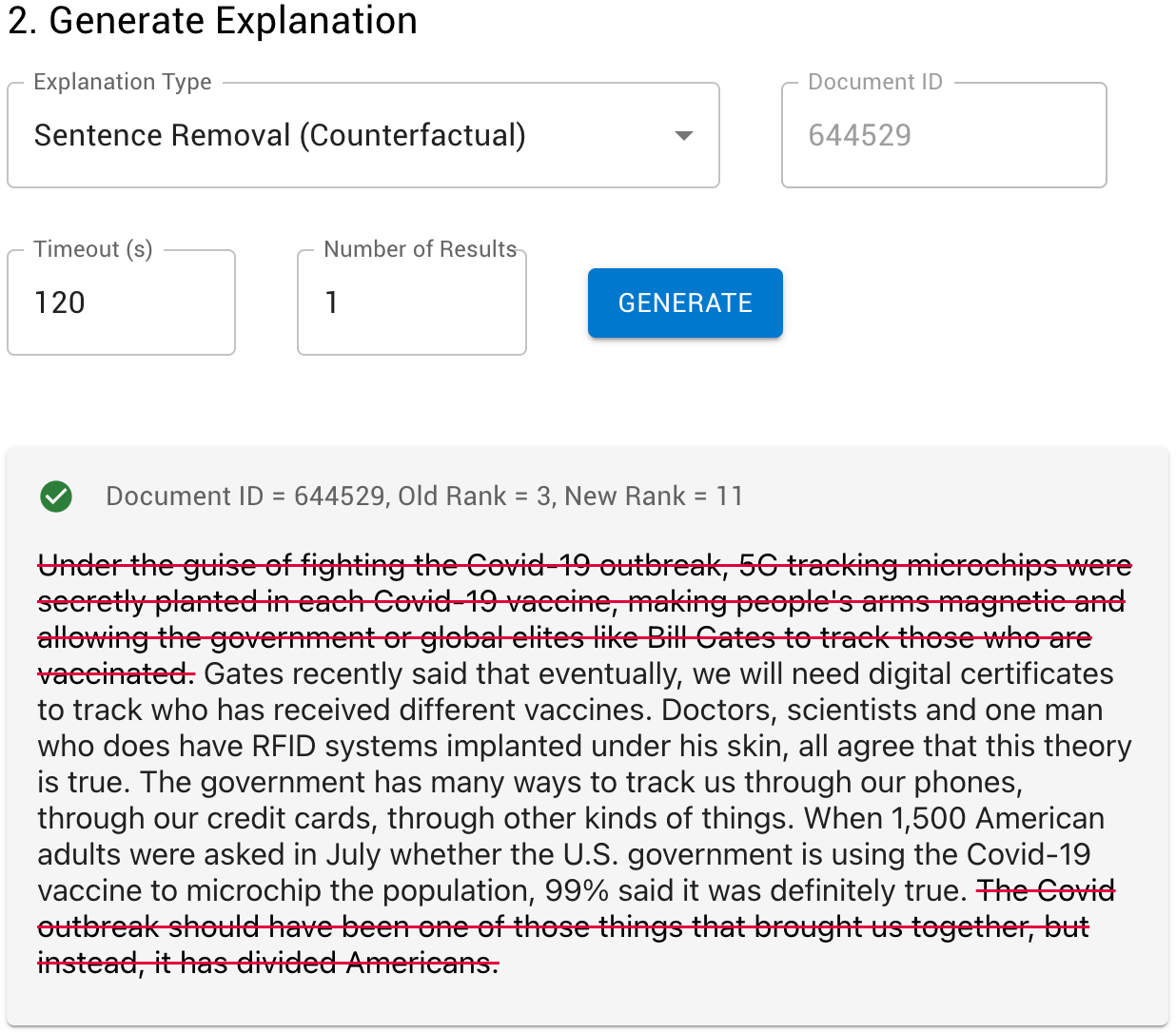}
  \vskip -0.1in
  \caption{
  A valid counterfactual perturbation
  for the selected document. 
  %If all
  %highlighted sentences are removed,
  %the rank of the document is lowered
  %from 3 to 11, which exceeds the
  %counterfactual threshold $k = 10$.
  }
  \label{documentexplfig}
\end{minipage}
\hfill % maximize horizontal separation
\begin{minipage}[t]{0.315\textwidth}
  \includegraphics[width=\linewidth]{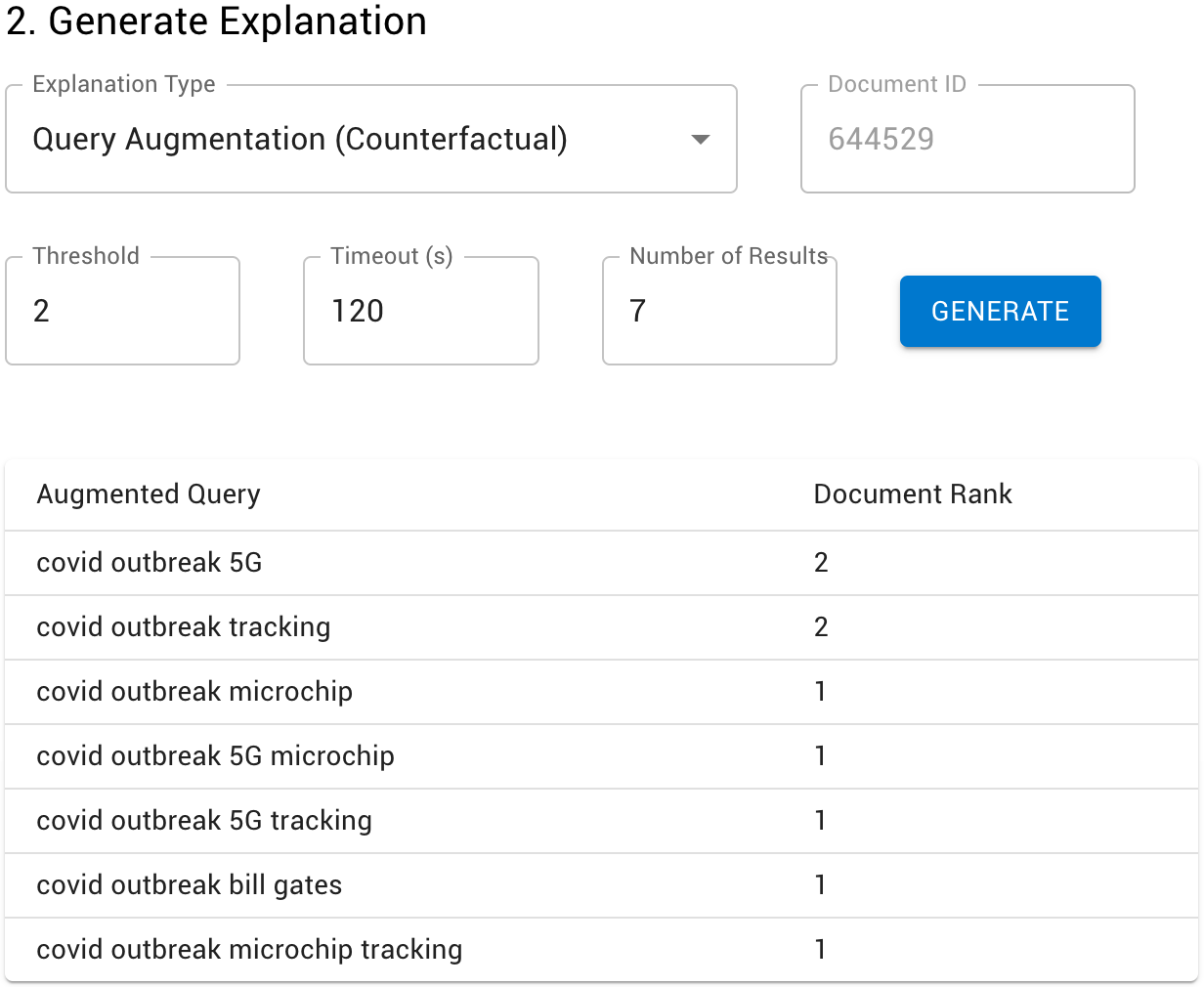}
  \vskip -0.1in
  \caption{
  Seven counterfactual query
  explanations 
  augmenting the original query
  ``covid outbreak''. 
  %Each query
  %results in the selected document
  %(not shown)
  %being given a rank $r' \leq 2$.
  }
  \label{queryexplfig}
\end{minipage}
\hfill % maximize horizontal separation
\begin{minipage}[t]{0.315\textwidth}
  \includegraphics[width=\linewidth]{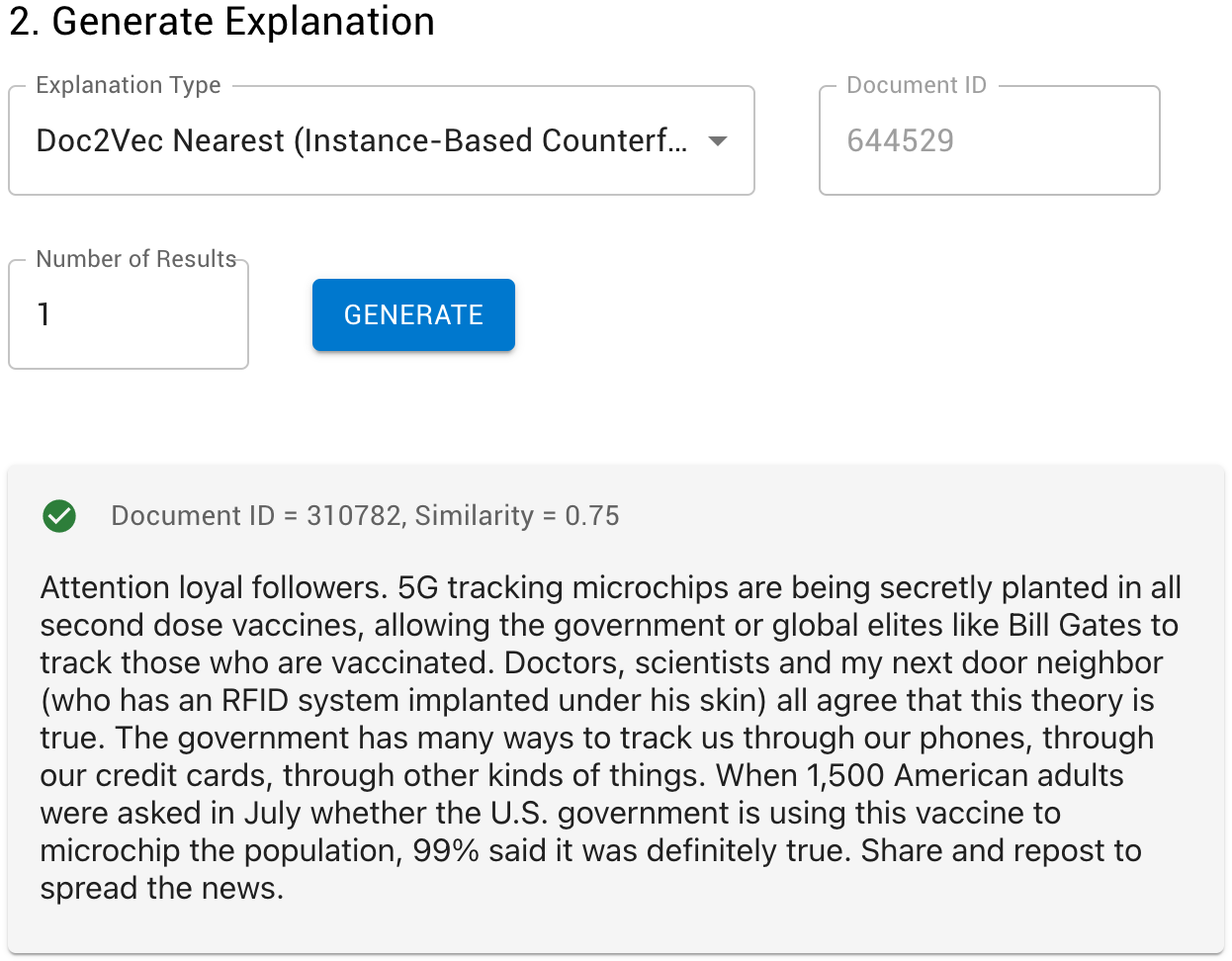}
  \vskip -0.13in
  \caption{
  A valid counterfactual document instance.
  %is displayed. This document is lexically
  %similar to the selected document
  %(not shown), however, it was not deemed
  %relevant.
  }
  \label{instancefig}
\end{minipage}
\vskip -0.14in
\end{figure*}

\subsection{Counterfactual Document Explanations}

To generate counterfactual explanations in
terms of a selected \textit{document} without corrupting its grammar,  
%one could perturb a document in many ways.
%Considering only removal-based perturbation,
%one might remove individual terms, phrases,
%sentences, or paragraphs. Using natural
%language processing (NLP) methods, such as
%topic modeling, or document statistics,
%such as TF-IDF, these document segments
%could be prioritized relatively,
%based on their
%semantic or lexical importance to the
%problem. However, \textit{removal} at
%certain granularities is much more
%likely to corrupt grammar or meaning,
%without additional correction.
%
%Without loss of generality, we
%implement a counterfactual algorithm that
%identifies relevance-negating document
%perturbations, formed only by 
we consider removing \textit{sentences}.
%In short, we believe that sentence removal
%is preferable for maintaining grammar and
%meaning in our tool. In our 
%counterfactual formulation, a 
An explanation identifies a minimal subset of
sentences in a given instance document 
whose removal lowers the rank of
the document beyond $k$. 

%The algorithm is given a query $q$ and
%instance document $d \in D$ (to generate an
%explanation for), along with an integer $k$
%such that $R(q, d, D, M) \leq k$.
%To impose limits on runtime, the user must
%also specify a maximum number of desired
%explanations, $n$.
%Let $S_d$ denote the set of sentences in
%a document $d$. The set of all valid
%counterfactual document perturbations
%(explanations) is defined as 
%$$P = \{d' \mid S_{d'} \subset S_d \wedge R(q, d', D', M) > k\},$$
%where $D'= D \cup \{d'\} \setminus \{d\}$.
%Rather than computing $P$ entirely, which
%may be computationally prohibitive,
%we design a simple greedy algorithm
%to efficiently identify $n$ valid
%explanations.

Intuitively, in any query-based retrieval
setting, the removal of search query terms
from a document is likely to lower document
rank, at least more than non-query terms.
Building on this intuition,
% we propose a greedy algorithm that
we propose an algorithm that
calculates an importance score for each
sentence in the instance document $d$, equal
to the number of sentence terms that
appear in the search query $q$. The algorithm
then iterates through 
explanations in sorted order.
Candidate documents are first sorted by
perturbation size (i.e., number of removed
sentences) in increasing order, then
by their importance score (i.e., the sum of
importance scores across removed sentences)
in decreasing order.
In each iteration, the perturbed document
is reranked, then added to a final
explanation set $P$ if deemed
non-relevant. This process continues
until $|P| = n$, 
%$P' \subseteq P$, until $|P'| = n$,
where $n$ is a maximum number of desired
explanations. 
This method 
%has the added benefit of
guarantees explanation \textit{minimality},
as all perturbations with $j$ removals
must be evaluated before those with
$j+1$.

\subsection{Counterfactual Query Explanations}

To generate counterfactual explanations in
terms of a \textit{search  query}, 
%one could
%perturb the query in many ways,
%perhaps by adding, removing, replacing,
%or rearranging its terms. Our goal is
%to reveal (to the user) new terms that
%\textit{distinguish} a selected
%document from other relevant documents.
%Therefore, we implemented a
%counterfactual algorithm that
%forms perturbations only by appending
we append terms from the instance document to the query, which
intuitively increases the document's relevance with
every addition. Although other terms
and other types of perturbation could be
used, they are likely to identify
relevance-raising search query perturbations
at a much slower pace. In our specific
formulation, a valid explanation
identifies a minimal
set of terms that, when
appended to the query, raises the rank of a
selected document beyond some threshold.

% Due to the large search space, we again
% propose a greedy algorithm 
Once more, we propose an iterative algorithm
%Identically to our document explanation
%formulation, this algorithm is also
%given a query $q$, instance document
%$d \in D$, and integers $k$ and $n$.
%% such that $R(q, d, D, M) \leq k$.
%Let $T_q$ denote the set of terms in
%some query $q$.
%The set of all valid counterfactual
%query perturbations is defined as 
%$$P = \{q' \mid T_{q} \subset T_{q'} \wedge R(q', d, D, M) \leq %k\}.$$
%Again, a simple greedy algorithm is
%employed 
%to test iteratively
to identify $n$ valid
explanations quickly.
%Similarly, let $T_d$ denote the set of
%terms in some document $d$.
Our algorithm builds a set of candidate
terms from the instance document,
%that do not appear in the search query, 
excluding terms that do not already
appear in the search query, 
%$C = \{t \mid t \in T_{d} \wedge t \notin T_{q} \}$,
% and aims to greedily evaluate terms
and aims to evaluate terms
in order of their importance to the
document. Although other
importance measures could be used,
we choose to score each candidate term
%$t \in C$ 
using TF-IDF, which scores terms based
on their frequency in, and exclusivity to,
the instance document $d$
(among the set of ranked documents $\boldsymbol{D}^M$).
All combinations of candidate terms 
%in $C$ 
are then iterated, first in
increasing order of perturbation size
(i.e., the number of appended terms),
then in decreasing order of their
TF-IDF scores (summed over
constituent terms).
As with our algorithm for
counterfactual document explanations,
iterating first by perturbation size
guarantees explanation minimality.

\subsection{Instance-Based Counterfactual Explanations}

To enable users to prioritize the
\textit{plausibility} of their
counterfactual
explanations, we implement two
instance-based (document)
counterfactual algorithms, which
output actual documents from the corpus
rather than arbitrary perturbations.
In our formulation, a valid
explanation for a relevant document
identifies a non-relevant document
with a high degree of similarity.
Here, \textit{relevance} is
dictated by $k$.

The instance-based algorithm is a
specialization of our regular
document counterfactual algorithm.
%, sharing congruent input and output
%formats. Given an instance
%document $d$ being explained,
%where $R(q, d, D, M) \leq k$
%for some $k$,
%the set of all valid counterfactual
%document perturbations is defined as 
%$$P = \{d' \mid d' \in D \wedge R(q, d', D, M) > k\}.$$
%Ideally, $d'$ must also be highly
%similar to $d$, by some measure.
To find a non-relevant document $d'$ that is similar to the
instance document $d$,
%Thus, 
we implement two variations of
the same counterfactual algorithm,
each employing different notions
of similarity and different document
sampling techniques. In the
first method, we train a Doc2Vec
embedding model \cite{le2014distributed}.
%an NLP tool that builds vector
%representations (i.e., embeddings)
%from a corpus, while directly
%encoding semantic similarity.
In the second method, we build
numeric vector representations
of each corpus document using
their BM25 scores, though any
similar collection statistic
(e.g., TF-IDF scores) would suffice.
In either case, with
numeric document vectors in hand,
we calculate similarity
using a cosine similarity
formula. In the first method,
we simply return
the $n$ most similar documents.
However, in the second method,
we sample $s$ non-relevant documents (ranked $k+1$ and below),
ideally where $n \ll s$,
then return the $n$ documents
with the highest similarity.
%In the next section, we present
%examples to illustrate this process
%and others.

\section{Demonstration Plan}

In this demonstration, conference participants will
generate minimal counterfactual
document and search query explanations,
instance-based document explanations,
and build their own document
explanations.
% Additionally, participants can
% apply topic modeling to quickly identify
% frequent term groupings across relevant
% documents.
Together, these components
enable diverse explainability for
individual ranking predictions.

\subsection{Counterfactual Document and Query Explanations}

\begin{figure*}[htbp]
\vskip -0.1in
\centerline{
    \includegraphics[width=0.9\textwidth]{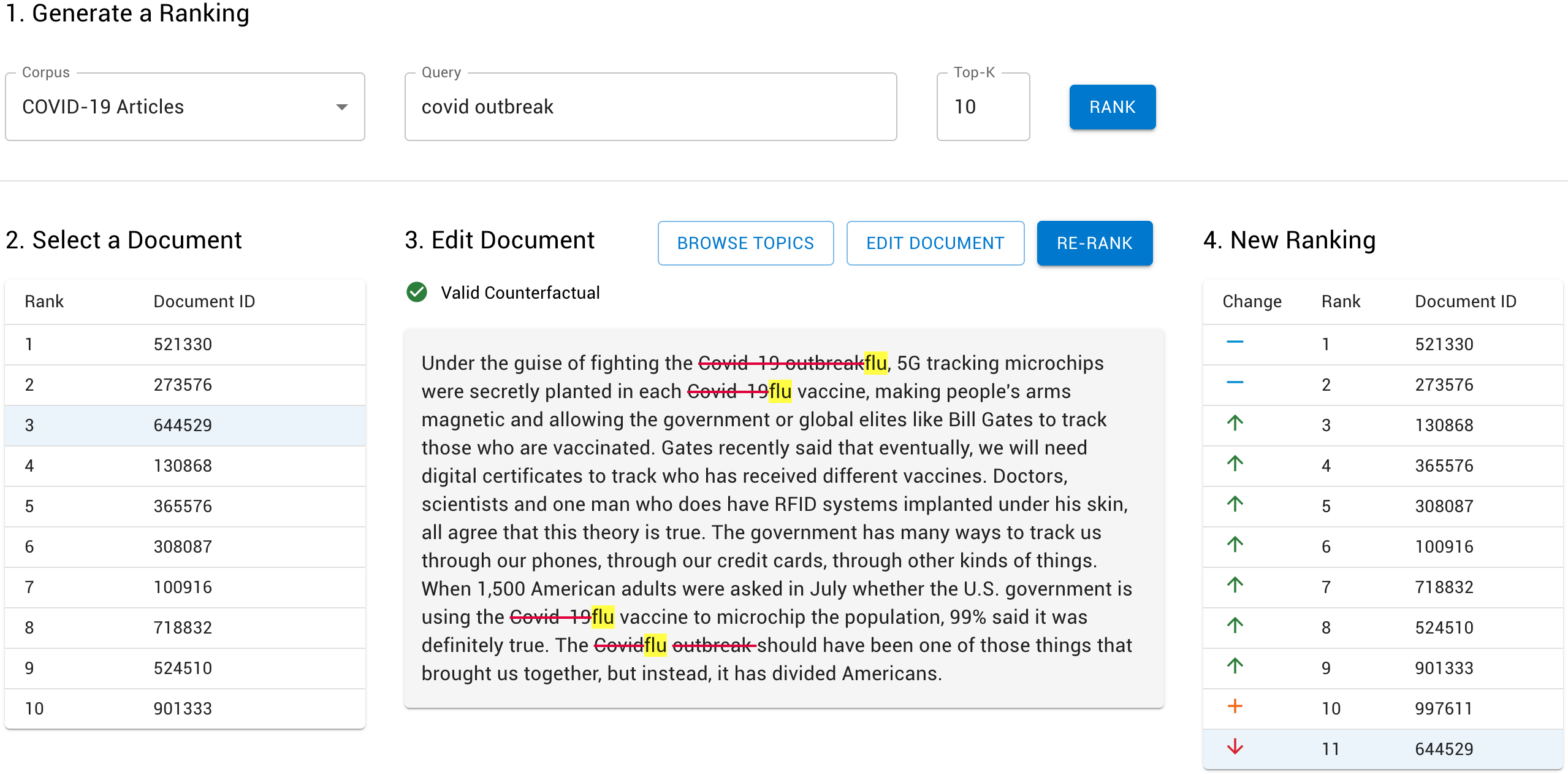}
}
\vskip -0.1in
\caption{
The counterfactual builder page. By replacing
all occurrences of `covid-19' with `flu', and
removing occurrences of `outbreak', the
document no longer ranks among the top-10.
The green check mark verifies this fact,
denoting the perturbation as a valid
counterfactual.
}
\vskip -0.13in
\label{builderfig}
\end{figure*}

On the \emph{Explanations} page,
the user is prompted to select a supported corpus,
type an arbitrary query, and select a
value of $k$. Once the
\keyword{Rank} button has
been clicked, a ranking of the top-$k$
documents appears beneath in a table.
By clicking individual documents in the table,
the user spawns a new \emph{Generate Explanation} pane
to the right,
from which four types of
counterfactual explanation may be
generated. In the following example,
we demonstrate and explain the
motivation behind the generation of
these explanations.

Consider a scenario where a user is
investigating a fake news (misleading information)
article that has ranked 3/10 in their
search for ``covid outbreak'', while
exploring the \emph{COVID-19 Articles} corpus.
Seeking document counterfactual
explanations,
the user selects the \emph{Sentence Removal}
type, requests one explanation, then clicks
\keyword{GENERATE}. As illustrated in Figure
\ref{documentexplfig}, the resulting
explanation renders the original body
of the document, crossing out
sentences that the counterfactual
perturbation has removed.
In this case, removing both sentences
mentioning \emph{covid} and \emph{outbreak}
lowers the document rank sufficiently
to render it non-relevant
(i.e., its rank of 11 surpasses $k = 10$).
Our algorithm, which scores
sentences by the number of query terms
present, assigns both the first and
last sentence a score of two. Thus,
they are heavily prioritized while
exploring perturbations, until their
combination (score of four) is discovered
to be a valid counterfactual.
Using this explanation, the user has
quickly learned why this fake news
article has ranked among the top-$k$.

Seeking to discover terms that
distinguish it from others,
the user now wonders which search queries
would raise the rank of this fake
news article even higher. With this
motivation, the user selects
the \emph{Query Augmentation} explanation
type, which generates a
\emph{search-query} counterfactual
explanation. Without changing their
query, the user selects this new
explanation type, and requests seven
explanations with a threshold of two.
A table of queries appears, seen in
Figure \ref{queryexplfig}. In this
case, the user learns that
the ranker would bestow the fake
news article a rank of 2/10 for
the augmented query
``covid outbreak 5G'', and
1/10 for
``covid outbreak 5G microchip''.
In our algorithm, these
distinguishing terms
(e.g., \emph{5G} and \emph{microchip}) are
assigned high (TF-IDF) scores,
since they do not appear in the
other nine relevant documents, and
therefore increase the priority
of query augmentations that
contain them.
By highlighting these terms,
these explanations yield insight
into the relevance of the
document within the corpus.
Moreover, the user may continue
reformulating their own search query,
perhaps using these insights
to discover other fake news articles.
% TODO: Discuss Minimality?

\subsection{Instance-Based Counterfactual Explanations}

On the \emph{Explanations} page,
two instance-based counterfactual methods
are available in the
\emph{Explanation Type} dropdown:
\emph{Cosine Sampled} and
\emph{Doc2Vec Nearest}.
The cosine
sampled explanation requires
a number of samples,
which controls the number of
documents for which the cosine
similarity is calculated.
In either case, each
resulting explanation is a
single document, whose body is
rendered beneath the prompt.

By evaluating the similarities and
differences between a
selected document
and counterfactual instance, a
user may gain insight
into the behavior of a ranker.
Continuing our example for the
query ``covid outbreak'',
the user selects 
\emph{Doc2Vec Nearest}
type from the dropdown. Upon
clicking \keyword{GENERATE}, a
valid counterfactual document
instance is rendered beneath the
prompt, stating its numeric
similarity to the document
being explained. The document
presented in the user's output
(Figure~\ref{instancefig})
is 75\% similar to the fake news
article being explained, despite
not being ranked among the original
top-10.

The inconsistency between a
document and its
counterfactual instance
inherently delineates a decision
boundary respected by the ranker.
Upon closer inspection of
Figure \ref{instancefig},
the user will notice that the
instance document is a near copy
of the original fake news article,
but likely ranked lower due to absence
of the terms \emph{covid} and \emph{outbreak}.
By exploring these instance-based
explanations, the user may discover
other fake news articles that were
absent from the original ranking,
while deriving insights about the
relevance of the original fake news
article. Moreover, presenting actual
instances bypasses the issues of
finding perturbations that maintain
grammar or meaning. In the next
subsection, we present one further
alternative to this perturbation issue:
allow the user to build
perturbations interactively.

% In other words, terms that appear
% in the selected document (relevant),
% but not in the counterfactual
% instance (non-relevant),
% jointly contribute to the
% top-$k$ relevance of the selected
% document. Terms appearing
% only in the counterfactual instance
% may explain its non-relevance,
% while terms common to both may
% have less influence on relevance.
% Moreover, instance-based
% explanations help users discover
% similar documents that were omitted
% under the original query and
% $k$ value. In this sense, users
% may also leverage counterfactual
% methodology to facilitate
% retrieval and exploration.

\subsection{Build-Your-Own Counterfactual Documents}

On the \emph{Builder} page,
users may build their own
counterfactual document perturbation,
then test its counterfactual validity
against the other ranked documents.
The user is prompted to select a
supported corpus, type an arbitrary
search query, and select a value of $k$.
Upon clicking the \keyword{RANK} button,
a ranking of the top-$k$ documents
is obtained from the ranking model,
and displayed inside a table.
% For
% the purposes of assessing
% counterfactual validity, we query
% the ranker for
% the top $k+1$ documents, but only
% show the top-$k$ in the table.
% Each
% row denotes a retrieved document, with
% rows sorted by rank (in decreasing
% order), and can be
% clicked to open the document in the
% counterfactual editor pane
% (ie. under `Edit Document').
Upon clicking a document in the
table, the document body is loaded
into an
interactive text field, allowing the
user to compose arbitrary edits.
The \keyword{BROWSE} \keyword{TOPICS}
button can be clicked to spawn a
modal, allowing the user to
generate and explore topics
found across all $k$ documents.
After finalizing document edits,
clicking the \keyword{RE}-\keyword{RANK} button
obtains a new ranking from the ranking
model.
Behind the scenes, the edited document
is substituted for the original, then
re-ranked alongside the other top
$k+1$ documents. The new ranking
of $k+1$ documents is displayed in
another table, with coloured
arrows to indicate whether the rank
of each document has been raised,
lowered, or left unchanged. The
originally hidden document with rank
$k+1$ is given an orange \emph{plus} icon
to distinguish itself.

In our running example,
the user poses the usual ``covid outbreak''
query for $k = 10$, then receives a familiar 
ranking of top-10 documents.
Clicking the fake news article at rank 3,
they 
%proceed to leverage their
%knowledge of the English language to
create several counterfactual
perturbations of their own. As
illustrated in Figure \ref{builderfig},
the user chooses to replace \emph{covid} and
\emph{covid-19} occurrences with an
alternative term \emph{flu}, and refactor the
term \emph{outbreak} in favour of
\emph{the flu}. After re-ranking, the
green check mark confirms
the counterfactual validity of the
perturbation, since its rank has
been lowered from 3 to 11 (i.e., $k+1$).
Using this interactive explanation
format, the user tested
their own plausible perturbations,
receiving valuable relevance insights
that transcend simple lexical
manipulations. In this example, the
user quickly learned how to edit
this fake news document,
so as to ensure it is not deemed
relevant to their query.

\bibliographystyle{IEEEtran}
\bibliography{IEEEabrv,references.bib}

\begin{thebibliography}{10}
\providecommand{\url}[1]{#1}
\csname url@samestyle\endcsname
\providecommand{\newblock}{\relax}
\providecommand{\bibinfo}[2]{#2}
\providecommand{\BIBentrySTDinterwordspacing}{\spaceskip=0pt\relax}
\providecommand{\BIBentryALTinterwordstretchfactor}{4}
\providecommand{\BIBentryALTinterwordspacing}{\spaceskip=\fontdimen2\font plus
\BIBentryALTinterwordstretchfactor\fontdimen3\font minus
  \fontdimen4\font\relax}
\providecommand{\BIBforeignlanguage}[2]{{%
\expandafter\ifx\csname l@#1\endcsname\relax
\typeout{** WARNING: IEEEtran.bst: No hyphenation pattern has been}%
\typeout{** loaded for the language `#1'. Using the pattern for}%
\typeout{** the default language instead.}%
\else
\language=\csname l@#1\endcsname
\fi
#2}}
\providecommand{\BIBdecl}{\relax}
\BIBdecl

\bibitem{poem}
V.~Dadvar, L.~Golab, and D.~Srivastava, ``Poem: Pattern-oriented explanations
  of {CNN} models,'' \emph{PVLDB}, vol.~15, no.~12, p. 3618–3621, 2022.

\bibitem{certa}
T.~Teofili, D.~Firmani, N.~Koudas, V.~Martello, P.~Merialdo, and D.~Srivastava,
  ``Effective explanations for entity resolution models,'' in \emph{{ICDE}},
  2022, pp. 2709--2721.

\bibitem{xaidb}
R.~Pradhan, A.~Lahiri, S.~Galhotra, and B.~Salimi, ``Explainable ai:
  Foundations, applications, opportunities for data management research,'' in
  \emph{SIGMOD}, 2022, p. 2452–2457.

\bibitem{explranking}
A.~Gale and A.~Marian, ``Explaining monotonic ranking functions,''
  \emph{PVLDB}, vol.~14, no.~4, p. 640–652, 2020.

\bibitem{singh2019exs}
J.~Singh and A.~Anand, ``Exs: Explainable search using local model agnostic
  interpretability,'' in \emph{WSDM}, 2019, pp. 770--773.

\bibitem{lirme}
M.~Verma and D.~Ganguly, ``Lirme: Locally interpretable ranking model
  explanation,'' in \emph{SIGIR}, 2019, pp. 1281--1284.

\bibitem{fernando2019study}
Z.~T. Fernando, J.~Singh, and A.~Anand, ``A study on the interpretability of
  neural retrieval models using deepshap,'' in \emph{SIGIR}, 2019, pp.
  1005--1008.

\bibitem{chios2021helping}
I.~Chios and S.~Verberne, ``Helping results assessment by adding explainable
  elements to the deep relevance matching model,'' \emph{arXiv preprint
  arXiv:2106.05147}, 2021.

\bibitem{Lin_etal_SIGIR2021_Pyserini}
J.~Lin, X.~Ma, S.-C. Lin, J.-H. Yang, R.~Pradeep, and R.~Nogueira,
  ``{Pyserini}: A {Python} toolkit for reproducible information retrieval
  research with sparse and dense representations,'' in \emph{SIGIR}, 2021, pp.
  2356--2362.

\bibitem{anserini}
P.~Yang, H.~Fang, and J.~Lin, ``Anserini: Enabling the use of lucene for
  information retrieval research,'' in \emph{SIGIR}, 2017, pp. 1253--1256.

\bibitem{lda}
D.~M. Blei, A.~Y. Ng, and M.~I. Jordan, ``Latent dirichlet allocation,''
  \emph{J. Mach. Learn. Res.}, vol.~3, pp. 993--1022, 2003.

\bibitem{le2014distributed}
Q.~Le and T.~Mikolov, ``Distributed representations of sentences and
  documents,'' in \emph{ICML}, 2014, pp. 1188--1196.

\end{thebibliography}

% \bibitem{b1} G. Eason, B. Noble, and I. N. Sneddon, ``On certain integrals of Lipschitz-Hankel type involving products of Bessel functions,'' Phil. Trans. Roy. Soc. London, vol. A247, pp. 529--551, April 1955.
% \bibitem{b2} J. Clerk Maxwell, A Treatise on Electricity and Magnetism, 3rd ed., vol. 2. Oxford: Clarendon, 1892, pp.68--73.
% \bibitem{b3} I. S. Jacobs and C. P. Bean, ``Fine particles, thin films and exchange anisotropy,'' in Magnetism, vol. III, G. T. Rado and H. Suhl, Eds. New York: Academic, 1963, pp. 271--350.
% \bibitem{b4} K. Elissa, ``Title of paper if known,'' unpublished.
% \bibitem{b5} R. Nicole, ``Title of paper with only first word capitalized,'' J. Name Stand. Abbrev., in press.
% \bibitem{b6} Y. Yorozu, M. Hirano, K. Oka, and Y. Tagawa, ``Electron spectroscopy studies on magneto-optical media and plastic substrate interface,'' IEEE Transl. J. Magn. Japan, vol. 2, pp. 740--741, August 1987 [Digests 9th Annual Conf. Magnetics Japan, p. 301, 1982].
% \bibitem{b7} M. Young, The Technical Writer's Handbook. Mill Valley, CA: University Science, 1989.
% \end{thebibliography}

\end{document}